\documentclass[sigconf]{acmart}
\usepackage{subfigure}
\usepackage{amsmath}
\usepackage{hhline}
\usepackage[table]{xcolor}
\usepackage{diagbox}
\usepackage{multirow}
\usepackage{makecell}
\usepackage{siunitx}

\AtBeginDocument{%
  }

\copyrightyear{2026}
\acmYear{2026}
\setcopyright{cc}
\setcctype{by}
\acmConference[ICONS 2026]{International Conference on Neuromorphic Systems}{August 04--06, 2026}{Chicago, IL, USA}
\acmBooktitle{International Conference on Neuromorphic Systems (ICONS 2026), August 04--06, 2026, Chicago, IL, USA}
\acmDOI{10.1145/3822454.3822476}
\acmISBN{979-8-4007-2808-2/2026/08}

\begin{document}

\title{A Neuromorphic Trigger for Efficient Audio Event Detection}

\author{Benjamin Hatton}
\orcid{0009-0007-5509-7873}
\affiliation{%
  \institution{ICNS, University of Manchester}
  \city{Manchester}
  \country{UK}
}

\author{Oliver Rhodes}
\orcid{0000-0003-1728-2828}
\affiliation{%
  \institution{ICNS, University of Manchester}
  \city{Manchester}
  \country{UK}
}

\author{Luca Peres}
\orcid{0000-0001-9748-9073}
\affiliation{%
  \institution{ICNS, University of Manchester}
  \city{Manchester}
  \country{UK}
}

\renewcommand{\shortauthors}{Hatton et al.}

\begin{abstract}
Efficient processing of continuous audio streams remains a key challenge for real-time and resource-constrained systems. This paper introduces a neuromorphic trigger for audio event detection, based on a spiking neural network (SNN) that selectively gates input to downstream models. The proposed neuromorphic trigger acts as a flexible low-cost front-end, identifying salient audio segments and enabling these to be processed by a more computationally intensive model for tasks such as classification.
The trigger is implemented as a lightweight fully connected SNN using a close-open filter for postprocessing, and is evaluated on two representative tasks: Anomalous Sound Detection (ASD) and Sound Event Detection (SED). For ASD, the trigger achieves a one-second segment-based F1 score of 0.97 on a class-agnostic form of the URBAN-SED dataset, demonstrating high reliability in identifying relevant audio regions. For SED, the trigger is combined with the Dang classifier on the DCASE 2017 Challenge Task 2 dataset, showing a potential $42.6\times$ reduction in FLOPs while reducing the lower bound of the event-based error rate from 0.41 to 0.25.
These results highlight the potential of neuromorphic triggers as real-time, energy-efficient front-end filters, enabling substantial reductions in computational cost.

\end{abstract}


\begin{teaserfigure}
  \centering
  \includegraphics[width=0.82\textwidth]{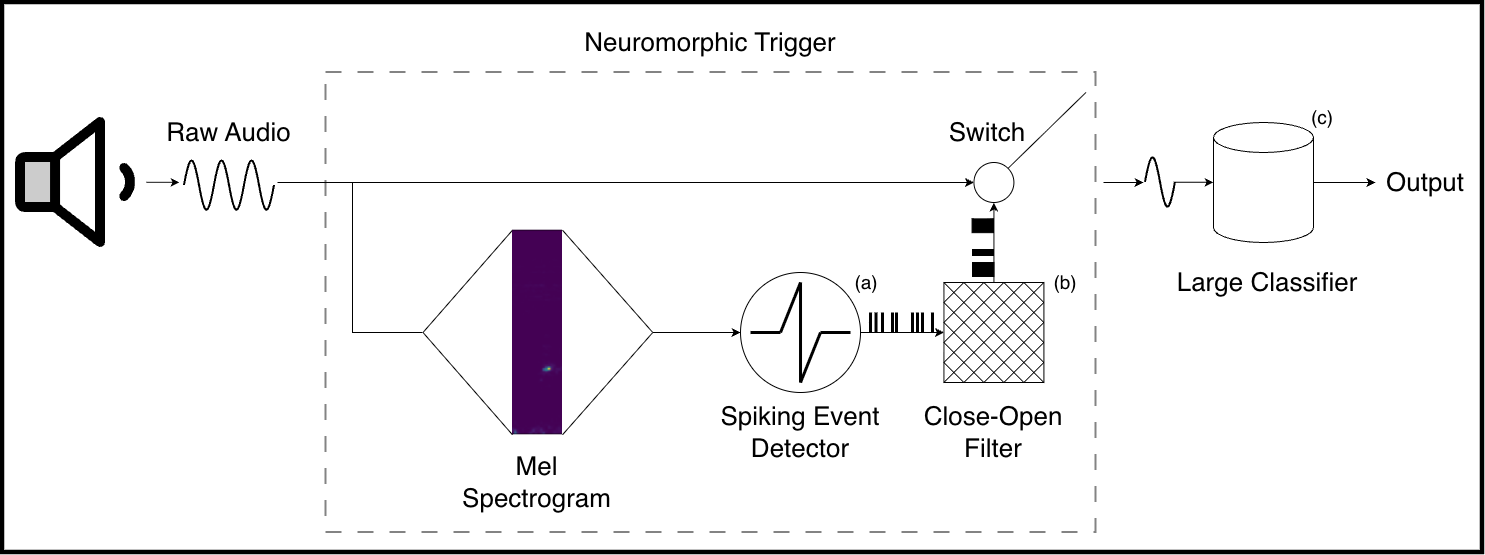}
  \caption{Proposed pipeline for more efficient processing of data. The audio is converted into a Mel spectrogram that is then passed to a spiking network (Spiking Event Detector). These spikes are then passed through a close-open filter to create contiguous blocks. The output of this filter triggers a switch that allows audio to be passed to a larger classifier.}
  \Description{Shows raw audio being processed into a Mel spectrogram, followed by a spiking network triggering a switch for the original audio to be passed to the large classifier.}
  \label{fig:teaser}
\end{teaserfigure}

\received{31 March 2026}
\received[accepted]{8 June 2026}
\received[revised]{24 June 2026}

\maketitle

\section{Introduction}
In a world where AI usage is ever expanding, methods for reducing the computational and power costs of such systems are growing in demand. Modern-day data processing is growing far more expensive than it has been in the past \cite{exp_costs}. This is due to the widening prevalence of transformer-based architectures, as well as the more intensive hardware required to run such models. With this in mind, recent research has attempted to find ways to circumvent these increased costs, and various approaches have been taken. For example, binary neural networks, spiking neural networks, and dynamic neural networks have all attempted to improve efficiency by simplifying the computations performed during inference.  
This work applies such principles to audio event detection. In this paper, audio event detection is used as an umbrella term to encompass two similar but distinct tasks: sound event detection (SED) and anomalous sound detection (ASD). 

Sound event detection is the problem of detecting and classifying certain patterns, known as events, within an audio sequence \cite{DCASE_og}. This can be viewed as target sounds occurring in environmental audio. For example, in the DCASE 2017 Challenge (Task 2), the task involved identifying gunshots, glass breaks, and babyies crying in a stream of background noise \cite{DCASE_2017}.
Most existing models for SED are designed to classify audio sequences into event classes while simultaneously identifying the onsets and offsets of the events. This combination of tasks forces the whole sample to be processed in its entirety, including the time between events of interest. This greatly increases the processing requirements compared with a model in which the data are filtered to retain only the events before classification.

The majority of existing architectures for sound event detection follow the same rough archetype. For example, convolutional recurrent neural networks (CRNNs) can be used for the detection of events in domestic environments \cite{CRNN2}. Alongside these traditional networks, gated recurrent units (GRUs) and long short-term memory (LSTM) layers are often used \cite{GRU_SED, LSTM_SED}. 
For more difficult tasks within SED, e.g. using heterogeneous data with missing labels, transformers are currently the state of the art \cite{Schmid2024}.

Anomalous sound detection (ASD) is the problem of determining whether some sound is normal or anomalous for a given environment \cite{asd_def}. A solution does not classify the sound, nor identify the cause. The state of the art for this problem includes audio transformers and autoencoders \cite{eat2025, auto2025}. 

Despite some attempts to reduce the power consumption of modern sound event and anomalous sound detection systems, e.g. with binary neural networks \cite{Binary_SED}, these networks still use expensive structures that integrate the data over time, requiring large amounts of memory and processing time. 

As spiking neural networks (SNNs) inherently integrate information over time, they are naturally suited to sequential-data domains \cite{SNN_3rd_Gen}. This makes them a suitable choice for audio-based tasks such as speech recognition \cite{SpikingHeidelberg}. Additionally, spiking networks have shown reductions in power consumption when applied to other auditory problems, e.g. keyword spotting \cite{neuro_key}. Given this --- and the pervasiveness of cloud computation --- a logical step towards efficiency is to reduce the amount of data being processed. 

To capitalise on this, we propose an SNN trigger model to be placed before a large classifier in the pipeline (as in Figure \ref{fig:teaser}). This model reads the audio input and detects sound events therein but does not attempt to classify them. The output represents a mask for the audio sequence, where a negative mask value means that the audio represented will not be processed. This has the potential to greatly reduce the required processing, as only the salient parts of the audio signal will be analysed.

\section{Related Works}

\subsection{Spiking Neural Networks}
Spiking neural networks (SNNs) are often posited as the third generation of neural networks \cite{SNN_3rd_Gen}. They are designed to function more like the brain, taking inspiration from biology, partially by introducing a temporal dimension to information processing. Rather than all neurons continuously outputting a real number, spiking neurons emit a binary spike depending on prior inputs. This allows spiking networks to seamlessly integrate over time without the use of expensive structures.

For example, the Leaky Integrate and Fire (LIF) model is a spiking neuron that consists of a membrane potential that changes over time and a spiking output based on thresholding \cite{Lapicque1907}. The original work details a continuous-time mathematical model; however, for a digital system, a discrete-time version is used. The snnTorch implementation of the LIF neuron is used in this work \cite{snntorch}.

For a given neuron, its membrane potential is represented by $U[t]$, and its output by $S[t]$ for timestep $t$. For a neuron to spike, $U$ must exceed the spiking threshold for that neuron, $U_{thr}$. Therefore, at a given time $t$, we obtain the Heaviside step function:
\begin{equation}
    S[t] = \begin{cases}
        1 & U[t-1] > U_{thr} \\
        0 & Otherwise
    \end{cases}
\end{equation}
The reset mechanism is how the membrane potential changes after a spike is emitted. The method used in this work is subtraction \cite{snntorch}:
\begin{equation}
    U[t] = \begin{cases}
        \beta U[t-1] + I_{in}[t] - U_{thr} & S[t] = 1 \\
        \beta U[t-1] + I_{in}[t] & Otherwise
    \end{cases}
\end{equation}
$I_{in}[t]$ represents the input to the neuron at time step $t$. The value $\beta$ is the decay rate of the neuron, hence its ``leaky'' nature. The decay dictates the ``memory'' of the neuron, with higher values retaining more information over longer periods of time.

As the LIF functionality hinges on a Heaviside step function, the spike generation is non-differentiable. Therefore, to train via backpropagation, a surrogate gradient is typically used \cite{SurrogateGradient}. Surrogate gradients approximate the error gradient during a backward pass by replacing the Heaviside function with a smooth gradient function, e.g. fast sigmoid or atan, enabling credit assignment for multilayered networks.

\subsection{Audio Event Detection Evaluation} \label{sec:sed_eval}
Within audio event detection, there are two main evaluation paradigms used to assess performance: event-based and segment-based \cite{main_metrics}. The event-based paradigm acts as a holistic approach, treating the predictions as block events. To evaluate the timing of events, there is often a margin (``collar'') in which the model can err without being penalised. Standards for the collar vary depending on the dataset, with 500\,ms being common for the tougher DCASE challenges \cite{DCASE_2017}.

The two paradigms are most commonly evaluated using the following metrics: the F1 score and the acoustic event error rate (AEER). Both of which are calculated by counting the true positives (TPs), false positives (FPs), and false negatives (FNs). For event-based metrics these are counted as:
\begin{itemize}
    \item \textbf{True Positive}: the model output has an overlapping event with the same label as the ground truth.
    \item \textbf{False Positive}: the model output has an event in the same time frame but no correlation with the label.
    \item \textbf{False Negative}: an event in the reference has no correlation with an event in the system's output.
\end{itemize}

F1 score is the harmonic mean of precision (the percentage of positive outputs that are true), $P$, and recall (the percentage of correct answers that are recognised), $R$. An F1 score of $1$ would indicate perfect recognition.

\begin{equation} 
    P = \frac{TP}{TP+FP}, \quad 
    R = \frac{TP}{TP+FN}, \quad 
    F_1 = \frac{2PR}{P+R}. 
\end{equation}


Acoustic event error rate (AEER or error rate) is calculated by summing the substitutions (S), deletions (D), and insertions (I) and then calculating the mean over the number of events in the sample (N) (as shown in Eqs. 4-7). As such, error rate is calculated using only false positives (FPs) and false negatives (FNs), so a perfect model would achieve an AEER of $0$. For any given sample $k \in K$:
\begin{align}
    S(k) &= min(FN(k), FP(k)) \label{eq:subs} \\
    D(k) &= max(0, FN(k) - FP(k)) \label{eq:dels} \\
    I(k) &= max(0, FP(k) - FN(k)) \label{eq:ins} \\
    AEER &= \frac{\sum^{K}_{k}\left[S(k) + D(k) + I(k)\right]}{\sum^{K}_{k}N(k)} \label{AEER_basic} \\
     &= \frac{\sum^{K}_{k}max\left( FN(k), FP(k) \right)}{\sum^{K}_{k}N(k)} \label{AEER_simp}
\end{align}

The segment-based paradigm is similar, except it divides the sample into adjacent sequences of length $n$. If there is a spike in the output, then that segment is labelled as positive by the model. This is then compared to a similarly divided ground truth to determine the above statistics.

\subsection{DCASE 2017 Task 2}
The DCASE Challenge 2017 Task 2 is chosen as one option to explore the neuromorphic trigger system due to its sparse event data and real-world similarities \cite{DCASE_2017}. The goal of the challenge is to identify the onset, offset, and class of events occurring within a sequence. The dataset used is the TUT Rare Sounds 2017 dataset, which is a set of monophonic audio event samples.

\begin{figure}
    \centering
    \includegraphics[width=\linewidth]{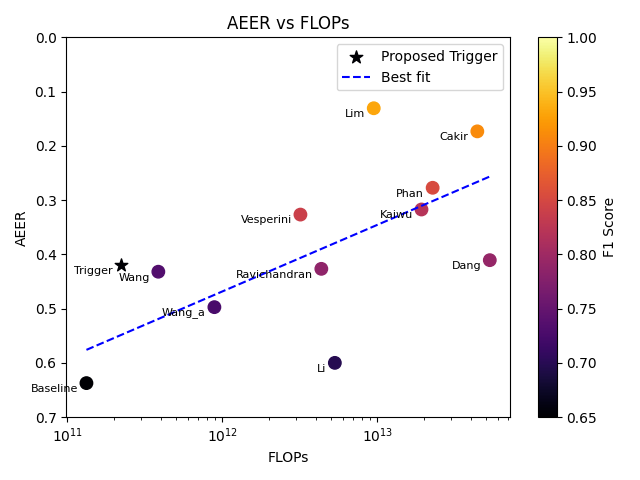}
    \caption{Comparison of the error rate (AEER) and the number of FLOPs of DCASE 2017 Challenge entries. The number of FLOPs is calculated using calflops \cite{calflops}, by passing through a 128 timestep sample, then using that to calculate FLOPs over the entire dataset. Also included is the trigger model introduced in Section \ref{nst} and evaluated in Section \ref{tut_res}.}
    \label{fig:er_comparison}
\end{figure}

The challenge solutions vary in size, method, and efficacy \cite{DCASE2017Results}. These are all then evaluated on a separate set from the dev-train and dev-test sets, resulting in two event-based metrics: F1 and error rate. For ranking the solutions, AEER is used over F1. Figure \ref{fig:er_comparison} shows the AEER of the network-based solutions against computational cost.

As can be seen in the figure, the method from \citet{Lim2017} performs best with an average computational cost, using a unidirectional LSTM with a convolutional backbone. This is then followed by \citet{Cakir2017} using a CRNN with a good AEER score of 0.1733 but a high FLOP count. The highest cost among the models is \citet{Dang2017} despite a relatively low AEER of 0.4107.

Within the challenge, many different approaches were employed. Some, like \citet{Phan2017}, attempt to teach background reduction as well as classification; while others, like \citet{Cakir2017}, use ensembles of models that have the same structure but different training. These approaches also varied in terms of architecture, including convolutional networks (CNNs) \cite{Dang2017}, variational autoencoders \cite{Ravichandran2017}, long short-term memory cells (LSTMs) \cite{Lim2017, Li2017}, and gated recurrent units (GRUs) \cite{Wang2017a}.


\section{Methods}
\subsection{Datasets} \label{sec:ds}
Within audio event detection, there are two forms of dataset: polyphonic and monophonic. A polyphonic set contains audio in which events are layered over background noise and possibly over other events as well. A monophonic dataset never overlaps two event sounds, creating a simpler classification task, although background noise may still be present.

Two datasets were used to evaluate the efficacy of the model. The first, URBAN-SED, is a synthetic dataset generated for polyphonic SED \cite{URBAN_SED_main}. The dataset consists of Brownian noise layered with various sound events, such that some sounds may occur simultaneously with others. URBAN-SED has ten classes, including ``dog bark'' and ``car horn''. The difficulty within this dataset is primarily in polyphonic recognition rather than in event detection itself due to the consistency of the background Brownian noise.

The second dataset explored is from the DCASE 2017 Challenge for Task 2 \cite{DCASE_2017}, the TUT Rare Sound Events 2017. This dataset is far more complex than URBAN-SED, despite being both monophonic and synthetic. The goal of the challenge was to generate datasets with real-world recordings; i.e. noise was recorded at various locations and overlaid with the events to be identified. The way in which the clips are overlaid often means that the events are very quiet and sunk into the background. This makes them difficult to identify, even for the human ear. On top of this, only $50\%$ of the samples contain an event, so no assumption about presence can be made; however, there is only ever at most one event within a sample.

The two datasets represent two different tasks in audio event detection. The TUT dataset, which has a noisy and variable background, requires sifting through the noise to try to find the recognisable sounds and is a prime example of SED. This could be applied to keyword spotting \cite{keyword} or identifying alarming noises in urban environments. On the other hand, URBAN-SED, due to its consistent background noise, represents the identification of abnormalities occurring in a relatively quiet, clean, and predictable environment. Therefore, it is a task far closer to anomalous sound detection (ASD) \cite{asd} due to the trigger being class-agnostic. This is an application more akin to wildlife monitoring or machine malfunction detection.

\subsection{Near-Sensor Trigger} \label{nst}
The proposed solution comprises a four-layer SNN made up of 384 LIF neurons (full structure and parameters in Table \ref{tab:geiger_structure}) and is situated at point (a) in the audio-processing pipeline (Figure \ref{fig:teaser}). This network is structured with a single output representing whether further processing is needed at a given timestep. The size of 384 neurons was selected to demonstrate the small scale possible, fitting within the TinyML standard of less than 1\,MB memory \cite{tinyml}. An SNN is used due to its potential low-power and low-latency when combined with neuromorphic hardware, further reducing computational costs. 

The output is used as a trigger for a large classifier, as seen in Figure \ref{fig:teaser}. The input of the trigger is a Mel spectrogram using 128 mel-bands for the TUT dataset and 64 mel-bands for the URBAN set. This spectrogram is created with a window of 64\,ms and a hop of 32\,ms, allowing for a 50\% overlap at every timestep. It was trained using the Adam optimiser with a learning rate of $0.001$ and a scheduler to reduce it on plateau (with a patience of $15$ epochs).

\begin{table}[]
    \centering
    \begin{tabular}{|c|c|} \hline
        Layer & Dimension \\ \hline
        1 & $64/128 \rightarrow 128$ \\
        2 & $128 \rightarrow 128$ \\
        3 & $128 \rightarrow 127$ \\
        4 & $127 \rightarrow 1$ \\ \hline
    \end{tabular}
    \begin{tabular}{|c|c|} \hline
         LIF Parameter & Value  \\ \hline
         Decay ($\beta$) & 0.8 \\ 
         Reset Mechanism & Subtraction \\
         Surrogate Gradient & Fast Sigmoid \\
         Threshold ($U_{thr}$) & $1.0$ \\ \hline
    \end{tabular}
    \caption{The structure and parameters of the proposed trigger network. The dimensions on the left detail the fully connected layers of LIF neurons.}
    \Description{A table displaying each of the 4 layers of the near-sensor spiking trigger network. Each layer is a set of fully-connected LIF neurons.}
    \label{tab:geiger_structure}
\end{table}

For training, a target output spike train is used. Both datasets supply on- and off-set times for each of the events, allowing timesteps situated between these two times to be labelled as positive. When comparing the output of the network to the target spike train, the Van Rossum distance is calculated \cite{VR_Distance}. This is then used as the loss for training the network through backpropagation, as in \cite{VR_Loss1, VR_Loss2}.

\begin{figure}
    \centering
    \includegraphics[width=0.9\linewidth]{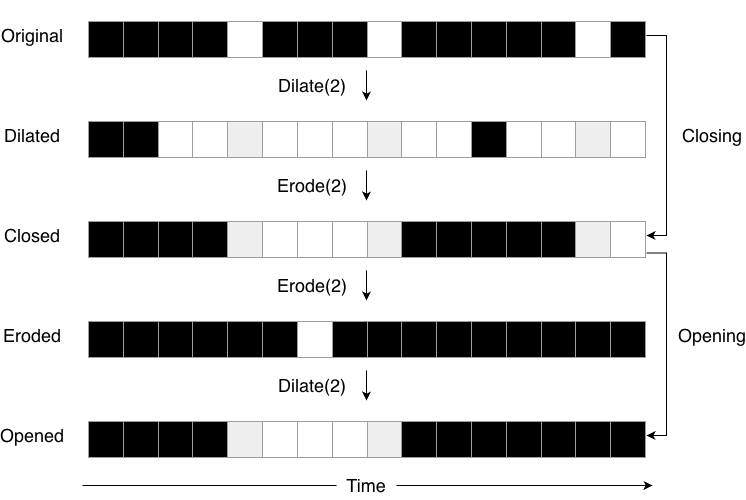}
    \caption{A diagram of the closing then the opening of a spike train modelled as a binary image. Here, white/black squares represent the presence/absence of spikes at a given timestep.}
    \Description{A diagram showing the closing then opening of the same spike train (modelled as a binary image) with each individual step labelled.}
    \label{fig:openclose}
\end{figure}

The output spike train of the model is then post-processed using a close-open filter (situated at (b) in Figure \ref{fig:teaser}), a concept from computer vision \cite{openclosebook}. Opening and closing are operations comprising erosion and dilation \cite{erodedilatebook}. Each of these is used on binary images to shrink and expand the size of the positive values within the image. Modelling the spike train as a one-dimensional binary image allows for the application of these processes, as seen in Figure \ref{fig:openclose}.

The post-processing more effectively defines the mask as it connects groups of discontiguous spikes (closing) and then removes any noise (opening). Any set of spikes that persists through the opening operation is seen as significant enough to be processed by the large classifier. The parameter $\xi$ dictates the size of the filter, e.g. Figure \ref{fig:openclose} shows $\xi = 2$.

\subsection{Post-Trigger Classifier}
Models for sound event detection (SED) classify the events, a post-trigger classifier must be used to enable direct comparison (situated at (c) in Figure \ref{fig:teaser}). To do so, the TUT dataset was chosen as it has the most benchmark results available in the literature. To show improvement, a model from the DCASE challenge was selected.  \citet{Dang2017} created a CNN- and GRU-based classifier with performance captured in Figure \ref{fig:er_comparison} (F1/AEER). This model was chosen due to its performance and because it has the most comprehensive technical report, enabling reproduction.

The Dang solution consists of three separate models, each acting as a binary classifier between its class and the other two. Within each model are two parallel pipelines, one comprising a CNN stack and the other a GRU. The two outputs are then aggregated and processed through a fully-connected (FC) layer to produce the model output. The three results are then concatenated and returned, with final prediction processing depending on the task.

The model is trained through backpropagation in two separate ways. One version treats each individual pipeline as its own model and uses binary cross-entropy loss. The other version treats it as a complete network, using cross-entropy loss (XE). As this is a conventional, non-spiking neural network, no surrogate gradients are needed. The standard Adam optimiser is used with a learning rate of $0.001$, and a scheduler that reduces the learning rate on plateau. As preprocessing, the data is cut to 100-timestep event samples using the supplied onset and offset to emulate the effect of the trigger.
If the selected region is less than 100 timesteps, the data is padded on both sides with data taken from the initial sample. If the event is longer than 100 timesteps, it is truncated.

The result of the model is a 3-tuple representing each class. As part of post-processing, the results are passed through a softmax function to turn the logits into a probability distribution \cite{softmax, softmax2}. After this, the 3-tuple is passed through a single FC layer to calculate a learned weighted sum of the output. The FC layer consists of three input neurons and three output neurons, acting as a linear transformation of the outputs to optimise the selection process.

\subsection{Evaluation}
The metrics used to quantify performance depend on the dataset. For the URBAN-SED dataset, one second segment-based scores are used as a standard, as seen in \cite{LEAD_DS_urban, YOHO}. The performance is measured using the F1 score, as this is the standard for ASD. Models typically found in the literature for this dataset predict class as well as onset and offset, e.g. the transformer-based architecture in \cite{urban_transformer}. 

As TUT Rare Sounds 2017 is a challenge dataset, there are several benchmarks to compare with. Evaluation for this dataset is done using error rate (AEER) rather than F1 as a standard, showing that mistakes are penalised more than correct predictions. We follow this trend as it allows for greater comparability between models. Again, the comparison models are classifiers that also predict onset and offset, rather than the separation of the tasks attempted here.

When evaluating the TUT set, the fact that there is at most one event per sample allows for the simplification of post-processing. After applying the close-open filter, the largest contiguous block of spikes can be taken from the output spike train to provide the prediction for the sample. 

Besides the statistical evaluation metrics, the number of Floating Point Operations (FLOPs) is counted and compared to those of other models. This is to show the reduction in processing requirements by using the neuromorphic trigger.

\subsubsection{Evaluation Challenges}
For a class-agnostic trigger, there are some challenges in terms of evaluation. Firstly, no information on similar models could be found; therefore, all evaluations must be done through comparisons to classifying models. This causes a discrepancy in the results, as the trigger does not focus on the classification of sounds but simply on detecting their presence.

As well as this, a spiking trigger can be incredibly sensitive. This is compounded by mistakes in the labelling of datasets. \citet{LEAD_DS_urban} have found that various annotators often label the data extremely differently, and this disparity can greatly affect the performance of detection models by upwards of $30\%$ on URBAN-SED.

\section{Results}
\subsection{Near-Sensor Trigger}
\subsubsection{URBAN-SED}
The model, when trained on URBAN-SED, achieved a segment-based F1 score of 0.9724 at $\xi=1$ (see Table \ref{tab:urban_results}) using the standard $1$\,s segment length. This shows that the trigger can effectively detect almost all events within the test set while not misidentifying the background noise.

A comparison of various filter sizes, $\xi$, and segment lengths can also be seen in Table \ref{tab:urban_results}. Within the table, it is clear that $\xi$ generally does not affect the results, except in two scenarios. First, having any $\xi > 0$ effectively removes errant spikes, as $\xi = 0$ allows noisy spikes to remain, decreasing the F1 score compared to other values of $\xi$. The second scenario is when $\xi$ almost exceeds or exceeds the length of the segment being analysed. This allows the output spike train to become more consistent by bridging gaps and, therefore, bridging the segments that would otherwise be empty.

The reason for the decrease in F1 at higher $\xi$ in Table \ref{tab:urban_results} is generally due to connections being made, sometimes spanning segments where they should not be. This causes a surge in false positives (reducing precision), dropping the F1 score. These surges are not as impactful as they could be because the higher $\xi$ also decreases false negatives (increasing recall), mostly mitigating the drop in F1. Therefore, the value of $\xi$ should be selected based on whether recall or precision is more important for the application.

The reduction in F1 as the segment resolution gets finer is expected, as the small discrepancies between the target and predicted spike trains become more prominent. The difference in F1 score between the $1$\,s segment and the $32$\,ms segment is low at 0.0215, showing that the model performs well at a range of resolutions with consistent output.

The models trained for SED on this dataset achieve a range of results. The You Only Hear Once model, based on the YOLO paradigm, achieved an average $1$\,s segment-based F1 score of 0.59 \cite{YOHO}. The transformer-based model achieved a score of 0.6577 \cite{urban_transformer} and a separate unspecified model achieved a score of 0.7276 \cite{LEAD_DS_urban}. Despite the model proposed here attaining a higher score, it is not directly comparable due to these results including classification.

\begin{table}[]
    \centering
    \begin{tabular}{|l||c|c|c|c|c|}  \hline
    \diagbox[width=\dimexpr \textwidth/9+0.85\tabcolsep\relax, height=0.75cm]{$\xi$}{Seg. Len.}
        & 1\,s & 0.5\,s & 0.25\,s & 0.1\,s & 32\,ms \\ \hhline{|=||=|=|=|=|=|}
        0 & 0.9718 & 0.9661 & 0.9590 & 0.9504 & 0.9455 \\ \hline
        1 (32\,ms) & \cellcolor{blue!15}\textbf{0.9724} & \cellcolor{blue!15}\textbf{0.9668} & \cellcolor{blue!15}\textbf{0.9592} & 0.9505 & 0.9500 \\ \hline
        2 (64\,ms) & 0.9718 & 0.9659 & 0.9588 & 0.9503 & 0.9505 \\ \hline
        3 (96\,ms) & 0.9713 & 0.9656 & 0.9588 & \cellcolor{blue!15}\textbf{0.9506} & 0.9508 \\ \hline
        4 (128\,ms) & 0.9708 & 0.9656 & 0.9589 & \cellcolor{blue!15}\textbf{0.9506} & \cellcolor{blue!15}\textbf{0.9509} \\ \hline
        5 (160\,ms) & 0.9705 & 0.9655 & 0.9589 & \cellcolor{blue!15}\textbf{0.9506} & \cellcolor{blue!15}\textbf{0.9509} \\ \hline
        6 (192\,ms) & 0.9702 & 0.9656 & 0.9591 & \cellcolor{blue!15}\textbf{0.9506} & 0.9508 \\ \hline
        7 (224\,ms) & 0.9696 & 0.9654 & 0.9590 & 0.9503 & 0.9504 \\ \hline
        8 (256\,ms) & 0.9686 & 0.9650 & 0.9585 & 0.9496 & 0.9497 \\ \hline
        9 (288\,ms) & 0.9682 & 0.9648 & 0.9580 & 0.9489 & 0.9489 \\ \hline
    \end{tabular}
    \caption{Comparison of the segment-based F1 scores of the URBAN-SED tests using various filter sizes, $\boldsymbol{\xi}$, and segment lengths. Note that 32\,ms is the length of one frame so represents a frame-level score.}
    \label{tab:urban_results}
\end{table}

\subsubsection{TUT Rare Sounds 2017} \label{tut_res}
For TUT Rare Sounds, the lowest event-based AEER achieved by the proposed model is 0.416 (with an F1 of $0.584$). This was attained without using the close-open filter (i.e. $\xi=0$). The best F1 attained was $0.617$ (with an AEER of $0.473$) using a size of $\xi = 5$ for the close-open filter (Table \ref{tab:expansion_results}). 

\begin{table}[]
    \centering
    \begin{tabular}{|>{\centering\arraybackslash}p{0.6cm}||c|c|c|c|c|c|c|} \hline
        $\xi$ & AEER & F1 & Prec. & Rec. & TPs & FPs & FNs \\ \hhline{|=||=|=|=|=|=|=|=|}
        0 & \cellcolor{blue!15}\textbf{0.416} & $0.594$ & 0.605 & \cellcolor{blue!15}\textbf{0.584} & 438 & 286 & 312 \\ \hline
        1 & $0.432$ & $0.611$ & 0.661 & 0.568 & 426 & 218 & 324 \\ \hline 
        2 & $0.441$ & $0.611$ & 0.675 & 0.558 & 419 & 201 & 331 \\ \hline 
        3 & $0.448$ & $0.611$ & 0.685 & 0.552 & 414 & 190 & 336 \\ \hline
        4 & $0.457$ & 0.612 & 0.701 & 0.542 & 407 & 173 & 343 \\ \hline 
        5 & $0.473$ & \cellcolor{blue!15}\textbf{0.617} & 0.746 & 0.526 & 395 & 134 & 355 \\ \hline 
        6 & $0.486$ & \cellcolor{blue!15}\textbf{0.617} & \cellcolor{blue!15}\textbf{0.774} & 0.513 & 385 & 112 & 365 \\ \hline
        7 & $0.509$ & $0.596$ & 0.761 & 0.490 & 368 & 115 & 382 \\ \hline 
        8 & $0.529$ & $0.578$ & 0.749 & 0.470 & 353 & 118 & 397 \\ \hline 
        9 & $0.562$ & $0.551$ & 0.745 & 0.437 & 328 & 112 & 422 \\ \hline
    \end{tabular}
    \caption{Results of the Trigger on the TUT Rare Sounds 2017 dataset with various filter size, $\boldsymbol{\xi}$. These comprise the Error Rate (AEER), Precision, Recall, True Positives (TPs), False Positives (FPs), and False Negatives (FNs).}
    \label{tab:expansion_results}
\end{table}

The full set of results for the different values of $\xi$ can be found in Table \ref{tab:expansion_results}. Figure \ref{fig:er_comparison} shows the performance of the trigger concerning the FLOP count of each of the other models tested. It must be noted, however, that this is solely the detection of an audio event rather than the classification of the sample.

The performance gained and lost through the close-open filter occurs due to the nature of the two metrics. 
Due to the simplification of the error rate in Eq. $8$, it can be seen that an increase in $\xi$ increases only one of these statistics (in this case, it is FNs, as seen in Table \ref{tab:expansion_results}). Comparatively, the F1 score is based on a ratio between precision and recall, and so a range of statistics affects it. It is also seen that lower $\xi$ is generally best for recall, as those that could be mistaken for erroneous spikes are not removed. Higher $\xi$ achieves greater precision but begins to miss the harder-to-identify events.

Table \ref{tab:classwise_tut} shows the class-wise detection results of the network. It is observed that the length of each sound greatly alters the performance of the filter on the resultant spike trains. As can be seen, the ``glass break'' class (an average of $0.79$ seconds long) is the easiest to detect for all tested values of $\xi$ except for the largest two. The ``gunshot'' ($0.86$\,s) is the hardest to detect and is most easily identified with no filter (i.e. $\xi=0$). The ``baby crying'' ($1.82$\,s) is difficult, but some improvement is shown at higher filter sizes, as expected. This is due to the spike trains closing the gap between cries in the sample. The gunshot and the glass breaking perform better at lower $\xi$ values as they are shorter, sharper sounds than the ``baby crying''. Seemingly, the longer the event being identified, the better higher $\xi$ performs; however, more event classes need to be analysed to draw a stronger conclusion.

\begin{table}[]
    \centering
    \begin{tabular}{|>{\centering\arraybackslash}p{0.5cm}||c|c|c|c|c|c|} \hline
         \multirow{2}{*}{\makecell[c]{$\xi$}} & \multicolumn{3}{c|}{AEER} & \multicolumn{3}{c|}{F1} \\ \hhline{|~||-|-|-|-|-|-|}
         & Baby & Glass & Gun & Baby & Glass & Gun \\ \hhline{|=||=|=|=|=|=|=|}
         0 & 0.504 & 0.364 & \cellcolor{blue!15}\textbf{0.500} & 0.524 & 0.713 & \cellcolor{blue!15}\textbf{0.530}\\ \hline
         1 & 0.492 & \cellcolor{blue!15}\textbf{0.276} & 0.552 & 0.558 & \cellcolor{blue!15}\textbf{0.739} & 0.517 \\ \hline
         2 & 0.480 & \cellcolor{blue!15}\textbf{0.276} & 0.568 & 0.575 & 0.731 & 0.510 \\ \hline
         3 & 0.480 & 0.296 & 0.568 & 0.584 & 0.721 & 0.513 \\ \hline
         4 & 0.464 & 0.328 & 0.580 & 0.604 & 0.707 & 0.510 \\ \hline
         5 & 0.460 & 0.352 & 0.608 & 0.627 & 0.709 & 0.500 \\ \hline
         6 & 0.464 & 0.372 & 0.624 & \cellcolor{blue!15}\textbf{0.636} & 0.708 & 0.490 \\ \hline
         7 & 0.464 & 0.408 & 0.656 & 0.627 & 0.685 & 0.459 \\ \hline
         8 & \cellcolor{blue!15}\textbf{0.456} & 0.448 & 0.684 & 0.634 & 0.650 & 0.429 \\ \hline
         9 & 0.460 & 0.504 & 0.724 & 0.635 & 0.606 & 0.387 \\ \hline
    \end{tabular}
    \caption{A table showing the results of the trigger system on the individual classes of TUT Rare Sounds 2017. The AEER and F1 are measured at various filter sizes, $\boldsymbol{\xi}$.}
    \label{tab:classwise_tut}
\end{table}

\subsubsection{Performance Evaluation} \label{sec:theoretical}
Assuming a perfect trigger system, theoretical computational costs can be calculated. The number of FLOPs that the trigger uses over the entire evaluation set is $F_{trigger}$. An ``event mask'', $E(t)$, represents the binary presence of an event at timestep $t \in T$, i.e. if an event is present at $t$ then the function returns $1$; otherwise, it returns $0$.

\begin{figure}
    \centering
    \includegraphics[width=\linewidth]{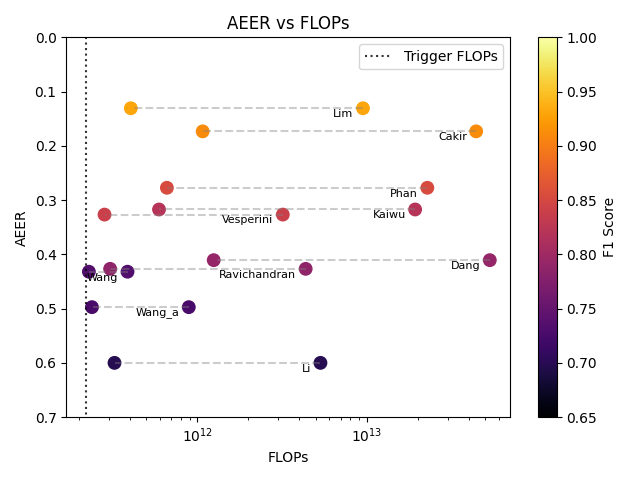}
    \caption{A comparison of the theoretical FLOP count assuming a perfect trigger system, calculated as in Section \ref{sec:theoretical}, with the original model. The grey lines represent the shift of each model, with the dotted black line representing the FLOP count of the trigger itself.}
    \label{fig:theoretical}
\end{figure}

For a given classifier, the number of FLOPs it uses per timestep is represented by the constant $C$ giving the equation:
\begin{equation}
    FLOPs=F_{trigger}+\sum^{T}_{t} \left( E(t) \cdot C \right) \label{eq:flop_count}
\end{equation}
Using Eq. \ref{eq:flop_count}, the theoretical lower bound can be estimated for each classifier, as shown in Figure \ref{fig:theoretical}. As can be seen, the potential improvement is directly proportional to the size of the classifier, with larger models using orders of magnitude fewer FLOPs for inference (up to $42.6\times$).

Alongside FLOPs, an estimate of the energy was calculated (Table \ref{tab:energy_ests}). This was done using two different sets of hardware to map the spiking and non-spiking operations within the model. SENeCA was used as the target neuromorphic platform \cite{seneca}, with multiply-accumulate (MAC) cost theoretically calculated from the given instruction-level data. AC/MAC operations were evaluated according to the neurobench framework approach, with ACs represented as synaptic operations (SOps) for SENeCA \cite{ac_eq_sop}.

Aside from this, the arithmetic cost on $45$\,nm hardware was calculated using the numbers from \citet{45nm} as cited in the literature \cite{sp_yolo}. These costs do not include reads, writes, or baseline power consumption and have formed a reference evaluation metric. For both forms of hardware, the energy cost for each operation type is calculated as:
\begin{equation}
    E_{usage} = N_{op}E_{op}
\end{equation}
Where $N_{op}$ is the number of operations and $E_{op}$ is the energy cost of said operation. While these number would require real-world testing and validation, they demonstrate the small fraction of energy consumed by the trigger relative to the system-level consumption.

\begin{table}[]
    \centering
    \begin{tabular}{|c|c||c|c|c|c|} \hline
        Hardware & Op. & Num. & pJ/Op & Energy & HW Total \\ \hhline{|=|=||=|=|=|=|}
        \multirow{2}{*}{\makecell[c]{SENeCA\cite{seneca}}} & SOp. & $6.5$M & $13.3$ & \SI{87.3}{\micro\joule} & \multirow{2}{*}{\makecell[c]{\SI{359.28}{\micro\joule}}} \\ \hhline{|~|-||-|-|-|~|}
        & MAC & $15.4$M & $17.7$ & \SI{272}{\micro\joule} & \\ \hhline{|-|-||-|-|-|-|}
        \multirow{2}{*}{\makecell[c]{45\,nm\cite{45nm}}} & MAC & $15.4$M & $4.6$ & \SI{70.7}{\micro\joule} & \multirow{2}{*}{\makecell[c]{\SI{76.599}{\micro\joule}}} \\ \hhline{|~|-||-|-|-|~|}
        & AC & $6.5$M & $0.9$ & \SI{5.91}{\micro\joule} & \\ \hline
    \end{tabular}
    \caption{A table showing the energy estimates of running the trigger network on two different forms of hardware, SENeCA and 45\,nm CMOS \cite{seneca, 45nm}. This is separated into the respective operations: synaptic operations (SOp), multiply-accumulates (MACs), and accumulates (ACs). Each of the figures is an average for each 30-second sample from the TUT 2017 dataset.}
    \label{tab:energy_ests}
\end{table}

\subsubsection{Comparison}
As stated in Section \ref{sec:ds}, the two datasets represent different tasks. The URBAN-SED set, with its background of Brownian noise, analyses anomalies layered in predictable noise (ASD). From the results, it can be seen that the proposed solution is effective for this style of problem.

In the TUT dataset, the trigger is effective for its size, but the room for improvement is clear. Given a larger model, it is likely that the trigger can learn to better detect the more hidden sounds. The trade-off, however, is that a larger size increases latency and computational costs.

From the data collected, it can be seen that the trigger is currently most effective for anomaly detection rather than sound events. Despite this, the results show great potential for SED in terms of reduction in computational cost.


\subsection{Post-Trigger Classifier} \label{sec:posttrig}
Table \ref{tab:posttrig_results} shows the accuracy achieved by the classifier for various training methods, with AEER and F1 score being tested using the outputs of the neuromorphic trigger. Note that the accuracy is calculated only with samples in which an event occurs.
As can be seen, the binary classifiers achieve accuracy above $90\%$, but when they are combined into a 3-class classification, the accuracy falls. There is negligible difference in accuracy when comparing the complete model with and without an FC layer attached. The AEER and F1 of the two models also differ insignificantly. When comparing the XE trained model, the parallel trained models perform better, with a five-percentage-point increase in accuracy. Despite this, the XE model still performs relatively similarly to the others in terms of AEER, likely due to the trigger system's performance on the dataset.

The entire pipeline performs worse on the task than most of the other models listed in the challenge (performing better than the Li and baseline models). This is due to the combination of inaccuracy in the model and the error rate of the trigger. In fact, using a perfect trigger allows the binary trained model without an output layer to achieve an error rate of 0.2573, which is better than the original Dang model (which attained 0.4107).

Using this theoretical trigger also drastically decreases the FLOP count by $42.6\times$. Not only can the trigger system increase the accuracy of a model in such a problem, but it can also greatly reduce the computational cost.  This shows that the separation of SED into sub-problems simplifies the task, allowing the individual parts to be solved more effectively.

\begin{table}[]
    \centering
    \begin{tabular}{|c||c|c||c|c|} \hline
        Method & Section & Accuracy & AEER & F1 \\ \hhline{|=||=|=||=|=|}
        & Baby & $92.73\%$ & \cellcolor{gray!40} & \cellcolor{gray!40} \\ \hhline{|~||-|-||~|~|}
        Parallel & Glass & $90.60\%$ & \cellcolor{gray!40} & \cellcolor{gray!40} \\ \hhline{|~||-|-||~|~|}
        Binary & Gunshot & $93.27\%$ & \cellcolor{gray!40} & \cellcolor{gray!40} \\ \hhline{|~||=|=||=|=|}
        Training & Total (FC) & $79.60\%$ & 0.608 & \cellcolor{blue!15}\textbf{0.454} \\ \hhline{|~||-|-||-|-|}
        & Total (Sep.) & $79.73\%$ & \cellcolor{blue!15}\textbf{0.596} & 0.433 \\ \hhline{|=||=|=||=|=|}
        XE Loss & Total & $74.80\%$ & 0.620 & 0.521 \\ \hline
    \end{tabular}
    \caption{Post-trigger classifier results. AEER and F1 represent TUT Rare Sounds 2017 scores using neuromorphic trigger outputs. Accuracies use the previously mentioned 100-timestep samples that contain events.}
    \label{tab:posttrig_results}
\end{table}

\section{Conclusion}
In this work, we present a method that can greatly reduce the computational costs of analysing audio through the use of a trigger. Through theoretical FLOP analysis, it is clear that the neuromorphic trigger significantly reduces the computational cost of inference, achieving up to $42.6\times$ for the tested Dang model.

As well as this, the ``ideal'' trigger displays the potential improvement that can be gained in terms of accuracy for classifier models, with the error rate of the classifier dropping from 0.41 to 0.25. The separation of tasks within sound event detection allows each model to focus on its individual sub-problem, decreasing the error rate.

Alongside this, it is evident that for anomalous sound detection (ASD), small trigger systems are effective at removing the majority of uninformative data from processing. The 0.97 F1 score achieved displays its ability to select salient sections for further analysis while rarely including background sections. This is further supported by the performance at smaller segment lengths than standard (e.g. the one-frame size of 32\,ms), indicating that the trigger operates effectively at high resolutions.


This initial investigation shows promising results and has limitations worth exploring. First, the applicability of the model can be tested beyond synthetic datasets to further validate the presented approach. Alternatively, an exploration of keyword spotting and human speech detection using the trigger could prove its viability in low-power accessibility systems.

Secondly, the current model focussed on a single structure and size, but different styles of triggers may also prove effective. Usually, ASD solutions are based on autoencoders. While not yet applied to ASD, spiking autoencoders can be created at a variety of scales with success \cite{sp_ae}. Therefore, an analysis of small-scale spiking autoencoders may show improvements in computational cost in ASD.


Finally, while this work shows potential for the simplified problem of classification after the identification of an event (in Section \ref{sec:posttrig}), further research into the topic is warranted. A potential avenue would include training a larger classifier to recognise when there is no event present. Such a network could then feed back to the trigger for better detection accuracy, also enabling online learning.

The proposed trigger establishes a novel, effective approach to reducing computational and power costs in audio processing while also demonstrating the potential to improve classification accuracy for SED. Its lightweight design makes it well-suited for edge deployment, enabling local, real-time processing.

\section{Funding}
This research was supported by the UK Government through the EPSRC Edgy Organism project (EP/Y030133/1).

\bibliographystyle{ACM-Reference-Format}
\bibliography{refs}

\end{document}